\documentclass{iopart}
\usepackage{amsfonts}
\usepackage{graphicx}

\begin{document}
\title[SU(3) Richardson-Gaudin models]{SU(3) Richardson-Gaudin models: three-level systems.}
\author{S Lerma H  and  B Errea}
\address{ Instituto de Estructura de la Materia, CSIC, Serrano
123, 28006 Madrid, Spain}

\ead{lerma@iem.cfmac.csic.es}

\begin{abstract}
 We present the  exact solution of the Richardson-Gaudin models
associated with  the SU(3) Lie algebra. The derivation is based on a
Gaudin algebra valid for any simple Lie algebra in the rational,
trigonometric and hyperbolic cases. For the rational case additional  cubic integrals of motion are obtained, whose number is added to  that of the quadratic ones to match, as required from the integrability condition,  the number of quantum degrees of freedom of the model. We discuss different SU(3)
physical representations and elucidate the meaning of the
parameters entering in the formalism.  By considering a bosonic
mapping limit of one of the SU(3) copies, we derive new integrable
models for three level systems  interacting with two bosons. These
models include  a generalized Tavis-Cummings model for three level
atoms interacting with two modes of the quantized electric field.
\end{abstract}
 \submitto{\JPA}
\section{Introduction}

The   Richardson-Gaudin (RG) exactly solvable  models
\cite{Duke1,Links3}
 can be traced back to  the exact solution of the BCS Hamiltonian
given by Richardson in the early 1960's \cite{Rich1} and to the
integrable spin model developed by Gaudin in the seventies
\cite{Gaudin1}.  They are diagonalizable by  Bethe Ansatz
techniques and are the so-called classical limit of two
dimensional  vertex models. We refer the reader to \cite{poghoss},
where  the connection between the RG models and the inhomogeneous
XXX vertex model with twisted boundary conditions is  established
in  detail for  the case of the rank-1 SU(2) algebra. Previously,
in \cite{Jurco}  it was shown that the solution of the Gaudin
models [those without linear term in the integrals of motion,
cf.(\ref{integ}) below] associated with more general Lie algebras,
can be used to get a solution of the corresponding Classical
Yang-Baxter Equations (CYBE).  This connection has been largely
exploited to diagonalize the RG models (see \cite{Links3,Links2}
for some examples). In practice this method consists of using the
already known solutions of the general-Lie-algebra Yang-Baxter
Equations (obtained by using the Quantum Inverse Scattering
method)   to  obtain the corresponding solution of  the RG models
by  taking the classical limit of the respective Bethe equations.
From  a  Conformal Field Theory context,  Asorey, Falceto and
Sierra   \cite{Sierra} obtained
 the integrals of motion and respective  eigenvalues of the rational RG models for any simple Lie algebra as a limiting case of the Chern-Simons theory.

Here we follow an alternative and more direct  approach to diagonalize the RG models.
Though essentially equivalent, it  differs in practice.  This approach, where no reference to the YBE is  necessary,  is similar to the presented by Ushveridze \cite{Ush} for the
rational case. The method is based  on the introduction of an infinite
dimensional algebra (the Gaudin algebra) associated with the
Lie-algebra of a simple group. By taking a Casimir-like operator
in the Gaudin algebra, one gets the transfer matrix of the YBE
approach and, consequently, a set of independent quadratic integrals of motion. In  \cite{Ortiz} this formalism was extended to  the
trigonometric  and hyperbolic cases for the rank-1 algebras SU(2)
and SU(1,1). In this contribution we extend the Gaudin algebra to the
trigonometric and hyperbolic cases for any simple Lie algebra.

  As an application of the method, we work out  in detail  the RG models associated with the  rank-2 SU(3) algebra.
 Most of the physical applications of the  RG models presented so far are based on
 the rank-1 algebras SU(2) or SU(1,1). They cover a wide variety of physical problems
ranging from pairing Hamiltonains for fermion or boson systems
\cite{Ami,Duke2}, to spin models or generalized Tavis-Cumming
models \cite{Duke3}. Recently,  some applications to physical
problems of RG models based on higher rank algebras     have been
published. The detailed derivation of the exact solution was
considered for the SO(5) algebra of proton-neutron pairing in
\cite{Links1} and with more generality including numerical
applications in \cite{Duke6}. In  \cite{Lerma} the RG model
based on the non-compact SO(3,2) algebra is used in the context of
the Interacting Boson Model II, which describes the interaction
between two different species of bosons ($\pi$ and $\nu$) for each
non-degenerated level. In \cite{Links2}  the SU(4) RG models are
used to describe the interaction between two superconducting
systems.  Here we discuss different physical scenarios of
potential interest, associated with the SU(3) RG models.  They
include   dipole-dipole interactions between three level atoms,
isospinorial pairing   and generalized Tavis-Cummings models of
three level atoms interacting with two different bosonic modes
\cite{TC3}.

Contrary to rank-1 RG models, in higher rank algebras the number
of  independent integrals of motion is not exhausted by those
obtained from the  quadratic Casimir-like operators. More
integrals of motion can be obtained by considering higher degree
operators. Unfortunately, as it was shown in \cite{Talal1},  the
Casimir-like analogy to obtain integrals of motions  is valid up
to degree-three Casimir operators. More general formulae are
needed to get integrals of motion of degree greater than three
\cite{Talal2}. Nevertheless, since the SU(3) algebra has two
independent Casimir operators of degree two and three
respectively, the Casimir-like analogy can be used to obtain the
complet set of integrals of motion. In this contribution  it is
verified that, for the rational SU(3) RG models,  the number of independent integrals of motion
coincides with the number of quantum degrees of freedom as defined
in \cite{Zhang}. For other higher-rank algebras, which have  at least quartic
Casimir operators, the more general formulae of \cite{Talal2}  are
needed to obtain the complete set of integrals of motion.

This paper is organized as follows. In section two, the general Gaudin algebra and quadratric Casimir-like operators  are introduced.  A particular realization of the Gaudin algebra in terms of a direct product of $L$  Lie algebra copies  and complex valuated functions is introduced. The conditions to be satisfied by these functions are established, which are a generalization of the Gaudin conditions of the SU(2) case.  Three particular solutions to the Gaudin conditions are found (rational, trigonometric and hyperbolic) and it is shown how to obtain the corresponding sets of quadratic integrals of motion from the Casimir-like Gaudin operator. Special interest is paid to the introduction of a linear term in the integrals of motion through the addition of a constant shift to the Cartan members of the Gaudin algebra. Explicit formulae for the quadratic  integrals of motion and their eigenvalues are given.
In section three, focusing on the SU(3) algebra, we apply the
formalism of section two  to obtain  closed expressions
 for the quadratic integrals of motion and their eigenvalues  in the
more general scenario of arbitrary SU(3) irreducible representations.
As discussed above,  in order to satisfy the condition of integrability more integrals of motion are needed, which must be at least cubic in the generators. For the rational version we present new sets of integrals of motion coming from the Casimir-like Gaudin operator of degree three, and  it is verified that the total number of quantum constants of motion coincides with the  number of quantum degrees of freedom.
 In section four, the physical
meaning of the variables appearing in the integrals of motion and
eigenvalues is discussed for  different SU(3) physical
realizations.  Additionally, the limit of infinite degeneracy for a copy of the
SU(3) algebra is used to get a family of integrals of motion
related to a generalized Tavis-Cummings model 
 for three
level atoms interacting with two species of bosonic excitations.
Conclusion are given in the last section.

\section{Richardson-Gaudin models from a Generalized Gaudin algebra}

Let us begin by considering a simple Lie algebra, expressed in
terms of its Cartan-Weyl decomposition \cite{CFThe,Wyb74}:
\begin{equation} [\mathbf{S}^a,\mathbf{S}^b]=\sum_c C^{ab}_c \mathbf{S}^c, \end{equation}
where the  non-zero structure constants  are given by:
\begin{eqnarray} C^{i \alpha}_\alpha=-C^{\alpha i}_\alpha
=\alpha^i,& \ \ \ \ C^{\alpha
-\alpha}_i=\frac{2}{|\alpha|^2}\alpha^i,&\ \ \ \  C^{\alpha
\beta}_{\alpha+\beta}
=N_{\alpha,\beta}\nonumber,\label{structure}\end{eqnarray} the
Latin index runs over the $r$ Cartan-Weyl members of the Cartan
subalgebra ($r=$rank of the group), and Greek indexes refer to the
roots. $\alpha^i$ denotes the component $i$ of the root $\alpha$.

 Associated with this algebra
we propose the following infinite dimensional Gaudin algebra:
\begin{equation} [\mathbf{S}^a(\lambda),\mathbf{S}^b(\mu)]=\sum_c
C^{ab}_{c}\{X_b(\lambda-\mu)\mathbf{S}^c(\lambda)-X_a^*(\lambda-\mu)\mathbf{S}^c(\mu)\},
    \label{GaudAl}\end{equation}
where $X_a(\lambda)$ are meromorphic  functions associated with the
member $a$ of the Gaudin algebra and $\lambda$ is a complex
variable.
 From these commutation rules, it can be proved that the
Casimir-like operators of the Gaudin algebra,
 \begin{equation}
\mathbf{K}(\lambda)\equiv
\sum_{ab}g_{ab}\mathbf{S}^a(\lambda)\mathbf{S}^b(\lambda),
\label{CasGaud}\end{equation}
 where $g_{ab}$ is the inverse of the Killing
form, commute among themselves:
 \begin{equation} [\mathbf{K}(\lambda),\mathbf{K}(\mu)]=0 \ \ \ \
\forall\  \lambda,  \  \mu . \label{transmat}\end{equation}
  A realization of the Gaudin algebra is given by:
   \begin{equation}
\mathbf{S}^a(\lambda)=\sum_{m=1}^L\mathbf{S}^{a}_{m}
X_a(z_m-\lambda) \label{gaudop}, \end{equation}
 where the index $m$ runs over  $L$ different  copies
of the Lie algebra,   $\mathbf{S}^a_m$ is a generator of the
$m$-th copy, and $z_m$ is a set of completely free real
parameters. The operators $\mathbf{S}^a(\lambda)$ act upon the
space $V_1\otimes ...\otimes V_L$, with $V_m$ an irreducible
representation (irrep) of the Lie algebra. Given this specific
realization, the commutation rules (\ref{GaudAl}) impose some
conditions on the functions $X_a(\lambda)$. All the functions
associated with the elements of the Cartan subalgebra are equal,
real valuated and antisymmetric: \begin{eqnarray}
X_i(\lambda)&\equiv&Z(\lambda)=Z^*(\lambda) \ \ \ \ \forall i\label{realz} \\
Z(\lambda)&=&-Z(-\lambda).\end{eqnarray} The reality condition on
the functions $ Z(\lambda)$ comes from  keeping the hermiticity of
the Cartan-subalgebra  in its  Gaudin counterpart
 [$\mathbf{S}^i(\lambda)=(\mathbf{S}^i(\lambda))^\dagger$].
 On the other hand, since
$\mathbf{S}^\alpha(\lambda)=(\mathbf{S}^{-\alpha}(\lambda))^\dagger$
the functions associated with the Gaudin ladder operators must satisfy:
\begin{equation} X_\alpha(\lambda)=X_{-\alpha}^* (\lambda) \qquad\forall
\alpha .\label{compleX}\end{equation} Additionally they
 have to be anti-hermitic:
 \begin{equation}
X_\alpha(\lambda)=-X_\alpha^*(-\lambda).
 \end{equation}
 The last   conditions that have to  be satisfied by the
functions $Z(\lambda)$ and $X_\alpha(\lambda)$  are  a generalized
version of the   $SU(2)$-Gaudin conditions:
 \begin{eqnarray}
\fl
 X_\alpha(\mu-\lambda)X_\alpha(\lambda)+Z(\lambda-\mu)X_\alpha(\mu)-Z(\lambda)X_\alpha(\mu)=0
 \quad & \forall\ \alpha \nonumber\\
\fl
X_{\beta}(\mu-\lambda)X_{\alpha+\beta}(\lambda)+X_{\alpha}(\lambda-\mu)X_{\alpha+\beta}(\mu)
 -X_{\alpha}(\lambda)X_{\beta}(\mu)=0 \quad & \forall \alpha,  \beta \quad  {\hbox  {such
 that}}\nonumber\\
 &\alpha+\beta {\hbox{ is a root.}} \label{gaud1}
\end{eqnarray}
 Using the structure constants, eqs. (\ref{structure}),
(\ref{realz}) and (\ref{compleX}),  and denoting the
Gaudin-operators by the more common notation
$\mathbf{S}^i(\lambda)=\mathbf{H}^i(\lambda)$ and
$\mathbf{S}^\alpha(\lambda)=\mathbf{E}^\alpha(\lambda)$, the
following form for the  Gaudin algebra is obtained:

\begin{eqnarray}
\fl [\mathbf{H}^i(\lambda),\mathbf{E}^\alpha(\mu)]&=&\alpha^i\{
X_\alpha(\lambda-\mu)\mathbf{E}^\alpha(\lambda)-Z(\lambda-\mu)\mathbf{E}^\alpha(\mu)\
\} \\
\fl [\mathbf{E}^\alpha(\lambda),\mathbf{E}^{-\alpha}(\mu)]
&=&\frac{2}{|\alpha|^2}X_{\alpha}^*(\lambda-\mu) \{\alpha\cdot
\mathbf{H}(\lambda)-\alpha\cdot \mathbf{H}(\mu)
\}\label{gaudro}\\
\fl [\mathbf{E}^\alpha(\lambda),\mathbf{E}^\beta(\mu)]&=&
N_{\alpha,\beta}\{
X_{\beta}(\lambda-\mu)\mathbf{E}^{\alpha+\beta}(\lambda)-X_{\alpha}^*(\lambda-\mu)\mathbf{E}^{\alpha+\beta}(\mu)\},
\end{eqnarray}
where $\alpha\cdot\mathbf{H}=\sum_i^r \alpha^i\mathbf{H}^i$. From
equation (\ref{gaudro}) it is clear that the Cartan subalgebra
members of the Gaudin algebra are defined up to a constant term,
which can be freely added  without altering the commutation
relationships: \begin{equation}
 \mathbf{H}^{i}(\lambda)=\sum_{m}\mathbf{H}^i_m Z(z_m-\lambda) + k^i\mathbf{1},
 \label{shift}
 \end{equation}
where $\mathbf{1}$ is the unity operator in $V_1\otimes...\otimes
V_L$, and $k^i$ are $r$  free real parameters.


In the rest of the paper, we will only consider  the case where
all the functions associated with the positive root operators are
equal ($X_\alpha(\lambda)=X(\lambda), \ \ \forall \alpha>0$) and
let for the future a more general discussion. In this case, three
solutions to  the Gaudin conditions (\ref{gaud1}) are:
\begin{itemize}
\item  The rational solution
 \begin{equation} X(z_n-z_m)= Z(z_n-z_m)=\frac{1}{z_n-z_m}
\label{solrat}
 \end{equation}
\item The  trigonometric solution\begin{equation}
X(z_n-z_m)=\frac{\exp[\mathbf{i}(z_n-z_m)]}{\sin(z_n-z_m)},\ \
Z(z_n-z_m)=\cot{(z_n-z_m)} \label{soltrig}\end{equation}

\item The hyperbolic solution\begin{equation}
X(z_n-z_m)=\frac{\exp[\mathbf{i}(z_n-z_m)]}{\sinh(z_n-z_m)},\ \
Z(z_n-z_m)=\coth{(z_n-z_m)}.\label{solhyp} \end{equation}
\end{itemize}

The Casimir-like operator defined in (\ref{CasGaud}), acts as a
generator of the quadratic RG integrals of motion.   From  (\ref{gaud1}) and their
solutions (\ref{solrat},\ref{soltrig},\ref{solhyp}), it is easy to
show that the operator $\mathbf{K}(\lambda)$ can be written as:
\begin{equation} \mathbf{K}(\lambda)=\sum_{m=1}^L
\frac{\mathbf{C}_m^{(2)}} {(\lambda-z_m)^2}-2\sum_{m=1}^L
\frac{\mathbf{R}_m}{\lambda-z_m}+...\label{expan} \
,\end{equation}
 where $\mathbf{C}_m^{(2)}$ is the Casimir of degree two of the $m$-th Lie algebra copy. The operators $\mathbf{R}_m$  read:
 \begin{eqnarray}
 \fl
\mathbf{R}_m =\mbox{\boldmath{$\xi$}}_m+
 \sum_{n (n\neq m)}^L &\left(  \sum_{\alpha >0}
 \frac{|\alpha|^2}{2}\left(X(z_n-z_m) \mathbf{E}^\alpha_m \mathbf{E}^{-\alpha}_n+ X^\ast(z_n-z_m)\mathbf{E}^{-\alpha}_m \mathbf{E}^{\alpha}_n\right)
\right.\nonumber\\
 &\quad\left. + Z(z_n-z_m) \sum_{i=1}^r \mathbf{H}_m^i \mathbf{H}_n^i
\right),\label{integ}
\end{eqnarray}
 where the linear term
$\mbox{\boldmath{$\xi$}}_m=\sum_{i=1}^r k^i \mathbf{H}_m^i$ is a
member of the m-th Cartan subalgebra coming from the constant term
shift introduced in (\ref{shift}). Note that the coefficients
$k^i$ do not have an index $m$, i.e.,  even if each
$\mbox{\boldmath{$\xi$}}_m$ belongs to a different copy of the Lie
algebra,  the same linear combination respect to the corresponding
basis of the Cartan subalgebra is assumed for all the copies.

The commutativity of the Casimir-like Gaudin operators
(\ref{transmat}), implies commutativity among the set of operators
$\mathbf{R}_m$: \begin{equation} [\mathbf{R}_m,\mathbf{R}_n]=0\ \
\ \forall\  n,m=1,...L .\end{equation}
The eigenfunctions and eigenvalues of the set of integrals of
motion $\mathbf{R}_m$ are obtained by considering a Bethe ansatz,
which is solely written  in terms of the lowering operators of the
Gaudin algebra and a set of parameters to be determined. This
calculation is performed in \cite{Ush} for the rational case.
Repeating the same steps for the trigonometric and hyperbolic
cases  is a straightforward but laborious task. It consists of
applying the Casimir-like Gaudin operator to the ansatz. This
application yields  one term proportional to the original ansatz
and  terms which are not proportional. From the former one,  one can
obtain the eigenvalues of $\mathbf{R}_m$ by performing a similar
expansion as in (\ref{expan}).  By imposing the annulation of all
the non proportional terms, one gets the equations that determine
the parameters entering in the ansatz. We present  the results.
The eigenvalues of the operators $\mathbf{R}_m$ are given by:
\begin{eqnarray}\fl r_m=\sum_{ab}^r\xi^a
F_{ab}\Lambda^b_m+\!\!\sum_{n\ (n\neq m) }^LZ(z_n-z_m)\ \
\Lambda_m \cdot F \cdot \Lambda_n
+\sum_{a=1}^r\sum_{k=1}^{M_a}\Lambda_m^a \frac{|\alpha_a|^2}{2}
Z(z_m-E^a_k),\nonumber\\
\label{evalues}
 \end{eqnarray} where
$\alpha_a$ are the simple roots of the algebra (there are as many
simple roots as the rank of the algebra).
 The $r$ coefficients $\xi^a$
are related to the parameters $k^i$ of the linear term through:
 \begin{equation} \mbox{\boldmath{$\xi$}}_m=\sum_i k^i
\mathbf{H}_{m}^i=\sum_{ab}\xi^a F_{ab}\mathbf{h}^b_m,
\label{xartan} \end{equation} where  $\mathbf{h}^b_m$ are the
members of the Chevalley  basis of the m-th Cartan subalgebra.
The Chevalley  basis is  defined as
$$\mathbf{h}^a_m=\frac{2}{|\alpha_a|^2}(\alpha_a\cdot\mathbf{H}_m),$$ where
$\alpha_a$ is one of the $r$ simple roots and $\mathbf{H}_m$ is
the already introduced Cartan-Weyl  basis of the Cartan
subalgebra. $F_{ab}$ is a $r\times r$ matrix called the quadratic
form of the algebra, which is related to  the Cartan matrix
through $F_{ab}=(A^{-1})_{ab} \frac{|\alpha_b|^2}{2}$. The Cartan
matrix codifies completely the structure of the algebra and is
defined in terms of the simple roots: $A_{ab}=
\frac{2}{|\alpha_a|^2} (\alpha_a,\alpha_b)$, where  the scalar
product of the roots is defined as: $(\alpha_a,\alpha_b)=\sum_i
\alpha_a^i \alpha_b^i$. $\Lambda_m^b$ are the Dynkin labels (the
eigenvalues of the Chevalley basis) in the highest weight state
of the $m$-th irreducible representation  $V_m$,  and the product
$\Lambda_m \cdot F \cdot \Lambda_n$ is equal to $\sum_{ab}
\Lambda_m^a F_{ab} \Lambda_n^b$. The variables $E_k^a$ entering in
 (\ref{evalues}) determine the common eigenfunctions of the
operators $\mathbf{R}_m$, and  are the solutions of the following
Richardson-Bethe equations:
\begin{equation} \fl\sum_{b=1}^r\sum_{k'=1}^{M_b} A_{ba} Z(E_{k'}^b-E_{k}^a) -
\sum_{m=1}^L  \Lambda_{m}^a Z(z_m-E_k^a)=\xi^a,\label{rbeq} \qquad
(k=1,...,M_a ). \end{equation}  The number of these parameters
$E_k^a$ is:
\begin{equation} M_a=\frac{2}{|\alpha_a|^2}\sum_{b=1}^r F_{ba}\left(\sum_m
\Lambda_m^b-\lambda_0^b\right), \label{limites}\end{equation}
where $\lambda_0^b$ are the eigenvalues of the overall operators
$\sum_{m}^L \mathbf{h}_m^b$ (the sum of the Chevalley  basis over
all the copies). These overall operators commute with the
integrals of motion $\mathbf{R}_m$ (see (\ref{noverall}) below),
therefore their eigenvalues ($\lambda_0^b$) are conserved
quantities for any Hamiltonian defined as a function of the
integrals of motion $\mathbf{R}_m$. Equations
(\ref{integ}),(\ref{evalues}), and (\ref{rbeq}) extend the results
presented in \cite{Sierra,Ush} for the rational case, to the
trigonometric and hyperbolic ones.

 \section{The SU(3) algebra}
\subsection{Quadratic integrals of motion}

In this section we will apply  the formulae of the previous one in
the particular case of the rank-2 algebra SU(3).   We  consider
$L$  copies of a $U(3)$ algebra:

\begin{equation} [\mathbf{K}_{\alpha\beta m },\mathbf{K}_{\alpha'\beta ' m}]=
\delta_{\beta\alpha'} \mathbf{K}_{\alpha\beta ' m}
-\delta_{\alpha\beta' }\mathbf{K}_{\alpha'\beta m}
\label{commutators},\end{equation} where the  index $m$ refers to
the $m$-th copy of the Lie algebra, and $\alpha,\beta=1,2,3$. From
these commutations rules it is easy to prove that $[\mathbf{K}_{11
m}+\mathbf{K}_{22 m}+\mathbf{K}_{33 m}, \mathbf{K}_{\alpha\beta
m}]=0 \ \ \forall \alpha,\beta$. Therefore this sum must be
proportional to the operator unity in the SU(3)-irrep $V_m$ and,
consequently, an integral of motion: \begin{equation}
\mathbf{K}_{11 m}+\mathbf{K}_{22 m}+\mathbf{K}_{33 m}\equiv
\mathbf{n}_m=n_m \mathbf{1}. \label{conserva}\end{equation}  This
condition reduces the number of independent generators from $9$ to
$8$, the dimension of the $SU(3)$-algebra. A  Cartan decomposition
of the SU(3) algebra is:
\begin{itemize}
\item
 A maximal abelian subalgebra of hermitian operators (the Cartan subalgebra
$\mathcal{H}$) is provided by the  operators $\mathbf{K}_{\alpha
\alpha m }$. However, due to the condition (\ref{conserva}) only
two of them are independent. Two different Cartan subalgebra bases
adequate for our purposes are the Cartan-Weyl basis (we are
using the normalization convention of  \cite{CFThe}):
\begin{equation} \mathbf{H}^1_m=\frac{\mathbf{K}_{11
m}-\mathbf{K}_{33 m}}{\sqrt{2}} \qquad
\mathbf{H}^2_m=\frac{2\mathbf{K}_{22 m }-\mathbf{K}_{33 m}
-\mathbf{K}_{11 m }}{\sqrt{6}},\label{ortogcar}\end{equation} and
the Chevalley  basis:
\begin{equation} \mathbf{h}_m^1=\mathbf{K}_{11 m}-\mathbf{K}_{22
m}\qquad
    \mathbf{h}_m^2=\mathbf{K}_{22 m}-\mathbf{K}_{33 m} \label{cheva}.
\end{equation} These bases generate the Cartan subalgebra
$\mathcal{H}_m\!\!=\!\!{\hbox { span}}(
\mathbf{H}^1_m,\mathbf{H}^2_m)\!\!=\!\!{\hbox{span}}(
\mathbf{h}_m^1,\mathbf{h}_m^2 )$.
 \item   The positive root vector space is spanned by the  raising operators:
\begin{equation}\mathcal{E}_m^+={\hbox{span}} (\mathbf{K}_{\alpha\beta m})
{\hbox{\ \ with\ \ }} \beta>\alpha. \label{rootp} \end{equation}
 \item The lowering
operators are the hermitian conjugated of the previous ones.  The
negative root vector space is: \begin{equation} \
\mathcal{E}_m^-={\hbox{span}} (\mathbf{K}_{\alpha\beta m})
{\hbox{\ \ with\ \ }} \beta<\alpha .\label{rootn}\end{equation}
\end{itemize}
The m-th Lie algebra of $SU(3)$  is given by the direct sum
$\mathcal{L}_{SU(3)
m}=\mathcal{E}_m^-\bigoplus\mathcal{H}_m\bigoplus\mathcal{E}_m^+$.
The roots of the algebra can be obtained from the commutation
relations between the Cartan-Weyl  basis  and the  root vectors:
\begin{equation} [\mathbf{H}_m^i,\mathbf{K}_{\alpha\beta
m}]=\alpha^i_{\alpha\beta} \mathbf{K}_{\alpha\beta m} \qquad
\alpha\not= \beta .\end{equation} The algebra SU(3) has  three
positive roots($\beta>\alpha$), they are:
\begin{eqnarray}
\alpha_{12} =& (\alpha^1_{12},\alpha^2_{12}) =&
\frac{1}{\sqrt{2}}(1,-\sqrt{3})\nonumber\\
\alpha_{23}=&(\alpha^1_{23},\alpha^2_{23})=&\frac{1}{\sqrt{2}}(1,\sqrt{3})\nonumber\\
 \alpha_{13} =&(\alpha^1_{13},\alpha^2_{13})=&(\sqrt{2},0)
\label{roots}.
 \end{eqnarray}
It is easy to see that the simple roots are $\alpha_{12}$ and
$\alpha_{23}$, from which all the other roots can be obtained by a
linear combination of integer coefficients. The non-simple
positive root is $\alpha_{13}=\alpha_{12}+\alpha_{23}$, whereas
the negative roots are: $\alpha_{21}=-\alpha_{12}$,
$\alpha_{32}=-\alpha_{23}$, and
$\alpha_{31}=-\alpha_{12}-\alpha_{23}$. All the roots have square
norm equal to 2: $|\alpha_{\alpha\beta}|^2=2$. The Cartan matrix
is: $
 A=\left( \begin{array}{cc}
    2 & -1  \\
    -1 &2  \\
\end{array}
    \right).
$
 From this expression we obtain the quadratic form:
$
 F=\frac{1}{3}\left(  \begin{array}{cc}
    2 & 1  \\
    1 & 2  \\
 \end{array}
 \right).
$

By using all these results, we can write the integrals of motion
(\ref{integ}) for the  SU(3) case. The rational
 integrals of motion read: \begin{equation} \fl\mathbf{R}_m= -\xi^1\mathbf{K}_{22 m}- (\xi^1+\xi^2)\mathbf{K}_{33 m} +C_m
\mathbf{1} +\sum_{n (n\neq m)}^L
 \frac{\sum_\beta\sum_\alpha\mathbf{K}_{\alpha \beta m}
\mathbf{K}_{\beta\alpha n}}{z_n-z_m} .
 \label{integsu3}\end{equation} For the trigonometric and
hyperbolic cases [${\hbox{ct}}(z_m-z_n)=\cot(z_m-z_n)$ and
${\hbox{ct}}(z_m-z_n)=\coth(z_m-z_n)$ respectively], the integrals
of motion are: \begin{eqnarray} \fl \mathbf{R}_m =&
-\xi^1\mathbf{K}_{22 m}- (\xi^1+\xi^2)\mathbf{K}_{33 m}
+C_m\mathbf{1} \label{integsu3tr} \\
\fl & +\!\!\! \sum_{n (n\neq m)}^L
\left[{\hbox{ct}}(z_n-z_m)\sum_\beta\sum_\alpha\mathbf{K}_{\alpha
\beta m} \mathbf{K}_{\beta\alpha n} + \mathbf{i}
\sum_{\beta>\alpha} (\mathbf{K}_{\alpha\beta
m}\mathbf{K}_{\beta\alpha n}-\mathbf{K}_{\beta\alpha
m}\mathbf{K}_{\alpha\beta n}) \right]\nonumber .
\end{eqnarray}
 The constant term appearing in (\ref{integsu3}) and (\ref{integsu3tr}) is: $C_m=-(n_m/3)\sum_{l (l\neq
m)} Z(z_l~-~z_m) n_{l}+
 \frac{(2\xi^1+\xi^2)}{3}n_m$.

\subsection{More integrals of motion}
By direct calculation, it can be shown that the overall operators
\begin{equation} \mathbf{N}_\alpha=\sum_ {m=1}^L \mathbf{K}_{\alpha\alpha
m},\qquad {\hbox{with}} \alpha=1,2,3,\label{noverall}\end{equation}
commute (in the trigonometric, hyperbolic,  and rational cases)
with the integrals $\mathbf{R}_m$.  One of these operators must be
independent of the already introduced integrals of motion. From
 (\ref{conserva}) it is clear that
$\mathbf{N}_1+\mathbf{N}_2+\mathbf{N}_3=\sum_m\mathbf{n}_m$, and
from the integrals of motion (\ref{integsu3}) or
(\ref{integsu3tr}) it is derived that
$\sum_m(C_m\mathbf{1}-\mathbf{R}_m)=\xi^1
\mathbf{N}_2+(\xi^1+\xi^2)\mathbf{N}_3$. This fact implies that
the integrals $\mathbf{R}_m$ do not exhaust all the possible
integrals of motion. In \cite{Talal1} it was shown for the
rational case that the Casimir-like Gaudin operators of degree
three commute among them and with the quadratic Casimir-like
operators  (\ref{CasGaud}). The SU(3) Casimir operator of degree
three is:
$\mathbf{C}^{(3)}=\sum_{\alpha\beta\gamma}\mathbf{K}_{\alpha\beta}\mathbf{K}_{\beta\gamma}\mathbf{K}_{\gamma\alpha}$,
from  here it is clear that the  Gaudin counterpart  must read:
 $\mathbf{K}^{(3)}(\lambda)=\sum_{\alpha\beta\gamma}\mathbf{K}_{\alpha\beta}(\lambda)\mathbf{K}_{\beta\gamma}(\lambda)\mathbf{K}_{\gamma\alpha}(\lambda)$,
with $\mathbf{K}_{\alpha\beta}(\lambda)$ a generator of the
SU(3)-Gaudin algebra. This family of  operators satisfies, at
least for the rational case,
$[\mathbf{K}^{(3)}(\lambda),\mathbf{K}^{(3)}(\mu)]=0$ and
$[\mathbf{K}^{(3)}(\lambda),\mathbf{K}(\mu)]=0$, $\forall \
\lambda, \ \mu$. By expanding  $\mathbf{K}^{(3)}(\lambda)$, as we
did for $\mathbf{K}(\lambda)$ in (\ref{expan}), we get:

 \begin{equation}
\fl
 \mathbf{K}^{(3)}(\lambda)=-\sum_{m=1}^L\frac{\mathbf{C}_m^{(3)}}{(\lambda-z_m)^3}+3\sum_{m=1}^L\frac{\mathbf{P}_{3m}}{(\lambda-z_m)^2}-3\sum_{m=1}^L
 \frac{\mathbf{R}_{3m}}{\lambda-z_m}+\sum_{\alpha=1}^3 A_\alpha \mathbf{1},
\end{equation}
where $\mathbf{C}^{(3)}_m$ is the cubic-Casimir of the $m$-th
Lie algebra copy, and the operators $\mathbf{P}_{3m}$ and
$\mathbf{R}_{3m}$ are:
 \begin{eqnarray}\fl
\mathbf{P}_{3m}=&\sum_{n(n\neq m)}^L\sum_{k {\mbox{\tiny
$\left(\!\!\!\begin{array}{c} k\neq
n\\
k\neq m\end{array}\!\!\!\!\right)$}}}^L\frac{
\sum_{\alpha\beta\gamma}
 \mathbf{K}_{\alpha \beta m}\mathbf{K}_{\beta\gamma n}\mathbf{K}_{\gamma \alpha
 k}}{(z_n-z_m)(z_k-z_m)}\nonumber\\
\fl
 &+\sum_{n(n\neq
 m)}^L\frac{\sum_{\alpha\beta\gamma}\mathbf{K}_{\alpha\beta n}\mathbf{K}_{\beta\gamma m}(\mathbf{K}_{\gamma\alpha n}-\mathbf{K}_{\gamma\alpha
 m})}{(z_m-z_n)^2}\nonumber\\
\fl
 &+\sum_{n (\neq
m)}\frac{\sum_{\alpha\beta}\left(A_{\alpha}\mathbf{K}_{\alpha\beta
m}\mathbf{K}_{\beta\alpha
 n}+A_{\alpha}\mathbf{K}_{\alpha\beta n}\mathbf{K}_{\beta\alpha
 m}\right) }{z_m-z_n}+ \sum_{\alpha}A_{\alpha}
A_{\alpha}\mathbf{K}_{\alpha\alpha m}\\
\fl \mathbf{R}_{3m}=&\sum_{n(n\neq
m)}^L\frac{\sum_{\alpha\beta\gamma}\mathbf{K}_{\alpha\beta m
}\mathbf{K}_{\beta\gamma m}\mathbf{K}_{\gamma\alpha n}}{z_m-z_n}+
\sum_{\alpha\beta}A_{\alpha}\mathbf{K}_{\alpha\beta
m}\mathbf{K}_{\beta\alpha m}.
\end{eqnarray}
The coefficients $A_\alpha$ are related to the parameters of the
linear term in $\mathbf{R}_{m}$: $A_1=(2\xi^1 +\xi^2)/3, \
A_2=(\xi^2-\xi^1)/3$ and $A_3=-(\xi^1+2\xi^2)/3$.
 The operators $\mathbf{P}_{3m}, \mathbf{R}_{3m}$ and $\mathbf{R}_{m}$ form a
complete set of mutually commuting operators. From the expression
for the integrals $\mathbf{P}_{3m}$, it is easy to show that the
three operators $\mathbf{N}_{\alpha}$ (\ref{noverall}) can be
expressed as a linear combination of the operators $\mathbf{R}_m$,
$\mathbf{n}_m$ and $\mathbf{P}_{3 m}$.  The commutativity of the
cubic Casimir-like Gaudin operators has been proved in
\cite{Talal1} for the rational case, we think this result can be
extended to the trigonometric and hyperbolic ones.

\subsection{Highest weight states and Quantum Dynamical Degrees of freedom}

 The highest weight states (unique for any
finite irrep of a simple algebra) are defined by the conditions:
\begin{equation} \mathbf{K}_{\alpha\beta m}|\Lambda_m\rangle=0
{\hbox{\ \ for all \ \ }} \beta>\alpha.
\end{equation} The eigenvalues of the members of the Chevalley
basis in these states are integer positive numbers and allow us to
label the irrep, \begin{eqnarray}
\mathbf{h}^1_m|\Lambda_m\rangle=\Lambda_m^1|\Lambda_m\rangle=(k_{1m}-k_{2m})|\Lambda_m\rangle  \nonumber\\
\mathbf{h}^2_m|\Lambda_m\rangle=\Lambda_m^2|\Lambda_m\rangle=(k_{2m}-k_{3m})|\Lambda_m\rangle
\label{labe},\end{eqnarray} where $k_{\alpha m}$ are the
eigenvalues of the operators $\mathbf{K}_{\alpha\alpha m}$ in the
highest weight state, and we have used (\ref{cheva}). The numbers
$[k_{1m}\ k_{2m}\ k_{3}]$ determine the Young diagram of the irrep
$V_m$ and satisfy: $ k_{1 m}+k_{2 m}+k_{3 m}=n_m$ . All the
members of the irrep can be obtained by acting the lowering
operators upon the previous highest weight state.

In general, once we have completely established  for each $m$ the
values of the set ($\Lambda_m^1,\Lambda_m^2,n_m$) (or equivalently
$[k_{1m}\ k_{2m} \ k_{3m}]$), we need three extra numbers for each
copy of the Lie algebra to determine a complete  basis  of the
quantum system. Consequently a complete basis for the Hilbert
space $V_1\otimes ...\otimes V_L$ is:
\begin{equation}|\Psi\rangle=\left |
\begin{array}{ccccc}
 n_1& ...&n_m &... &n_L\\
    (\Lambda^1_1,\Lambda^2_1)&...&(\Lambda^1_m,\Lambda^2_m)&...&(\Lambda^1_L,\Lambda^2_L)\\
    t_{11},t_{21},t_{31}&...&t_{1m},t_{2m},t_{3m}&...&t_{1L},t_{2L},t_{3L}
\end{array}\right\rangle_\rho
\label{basota} \end{equation} where $\rho$  is a multiplicity
number to take account of  any other completely degenerated
quantum number not considered explicitly here. The labels $t_{1m
},t_{2m},t_{3m}$ are the three necessary quantum numbers  to
characterize completely a state within a SU(3) irrep.
 The  number of  non completely degenerated quantum numbers necessary to determine
 unambiguously a basis' member is  the number of  quantum dynamical degrees of
freedom of the system \cite{Zhang}. From (\ref{basota})  it is
deduced that, in the present case, the number of quantum degrees
of freedom is:
 \begin{equation}
  d=3L.
\end{equation} To guarantee the integrability of the system we need $3L$
 integrals of motion. $L$ of them are provided by the
operators $\mathbf{R}_m$,   the other $2L$ integrals of motion are
the polynomial operators of degree three ($\mathbf{R}_{3m}$ and
$\mathbf{P}_{3m}$) introduced in the previous subsection.

\subsection{Eigenvalues}

 We can explicitly write the eigenvalues of the operators
(\ref{integsu3}) and (\ref{integsu3tr}), which are the SU(3)
version of the  general formula (\ref{evalues}):
\begin{eqnarray} \fl r_m=& -\xi^1 k_{2 m}-(\xi^1+\xi^2)k_{3 m}+ C_m +\sum_{n (n\neq m)}^L
\left(Z(z_n-z_m)\sum_{\alpha=1}^3 k_{\alpha m} k_{\alpha
n}\right)\nonumber\\
\fl & + (k_{1 m}-k_{2 m})\sum_{k=1}^{M_1} Z(z_m-E_k)+(k_{2 m}-k_{3
m})\sum_{l=1}^{M_2} Z(z_m-\omega_l),  \label{evaluesu3}
\end{eqnarray} where $C_m$ is the same constant found in the
operators (\ref{integsu3}) and (\ref{integsu3tr}), and we have
redefined the
 parameters $E_k\equiv E_k^1$ and $\omega_l\equiv E_l^2$. These
parameters determine the eigenfunction of the operators
$\mathbf{R}_m$, and  are  the solutions of the following
Richardson-Bethe equations:

\begin{eqnarray}
\fl
\sum_{k' (k'\neq k)}^{M_1} 2
Z(E_{k'}-E_k)-\sum_{l=1}^{M_2}Z(\omega_l-E_k)
-\sum_{m=1}^L (k_{1 m}-k_{2 m}) Z(z_m-E_k)&=\xi^1 \label{richeqsu3-a}\\
\fl
 & (k=1,.., M_1)\nonumber\\
\fl -\sum_{k=1} ^{M_1}Z(E_k-\omega_l)+\sum_{l'(l'\neq l)}^{M_2} 2
Z(\omega_{l'}-\omega_l)-\sum_{m=1}^L (k_{2 m}-k_{3 m}
)Z(z_m-\omega_l)& =\xi^2   \label{richeqsu3-b}\\
 & (l=1,.., M_2)\nonumber .
\end{eqnarray} To determine the number of parameters  $E_k$ and $\omega_l$
($M_1$ and $M_2$ respectively), note that the members of the
overall Chevalley basis  can be expressed in terms of the
integrals (\ref{noverall}),
$$\sum_m\mathbf{h}_{m}^1=\mathbf{N}_{1}-\mathbf{N}_2,\qquad
\sum_m\mathbf{h}_{m}^2=\mathbf{N}_2-\mathbf{N}_3,$$ consequently
 their eigenvalues  are: $(\lambda_0^1, \lambda_0^2)= (N_{1}-N_{2},
 N_{2}-N_{3})$. By using this result, the relation $\sum_{m}(k_{1m}+k_{2m}+k_{3m})=N_1+N_2+N_3$, the quadratic
form, and  the labels of the SU(3) irreps (\ref{labe}) in the
general formula (\ref{limites}), one gets: \begin{equation}
M_1=\sum_{m}^L k_{1 m}-N_1 \qquad  M_2=N_3-\sum_{m}^L k_{3
m}.\label{limitesu3} \end{equation}

The values $M_1$ and $M_2$ allow us to label the different
invariant subspaces of the   Hilbert space. We denote these
subspaces by $V(M_1,M_2)\subset V_1\otimes...\otimes V_L$.
 The different solutions of the Richardson-Bethe equations define a
set of eigenfunctions which span entirely the subspace
$V(M_1,M_2)$. By considering  all the possible values of $M_1$ and
$M_2$ and the set of complete solutions of the corresponding
Richardson-Bethe equations, we get a basis for the entire Hilbert
space $V_1\otimes ...\otimes V_L=\bigoplus_{M_1,M_2} V(M_1,M_2)$.

For a given set of $L$ SU(3) irreps ( $[k_{1m}\ k_{2m}\ k_{3m}]$ with $m=1,...,L$),  the
possible values of $M_1$ and $M_2$ are:
\begin{eqnarray}
\qquad\qquad\qquad\qquad\qquad 0&\leq  M_1  \leq & A_{13} \nonumber\\
\qquad\qquad \max\left(0,M_1-A_{12}\right)&\leq M_2 \leq &
A_{13}+ \min\left(0,M_1-A_{12}\right) \nonumber,
\end{eqnarray}
with $A_{13}=\sum_m (k_{1m}-k_{3m})$ and  $A_{12}=\sum_m (k_{1m}-k_{2m})$.

\section{Physical models related to the $SU(3)$ RG models }

\subsection{Particle-hole representation}
An explicit realization of the U(3) algebra (\ref{commutators})
can be obtained by  considering  $\Omega_m$ three-level atoms of
type $m$:

\begin{equation} \mathbf{K}_{\alpha\beta m
}=\sum_{\mu=1}^{\Omega_m}\mathbf{c}_{\alpha\mu m}^{ \dagger}
\mathbf{c}_{\beta\mu m}, \label{operators1}\end{equation} where
$\mathbf{c}_{\alpha\mu m}^{ \dagger}$ and $\mathbf{c}_{\beta\mu m
}$ are fermion creation and annihilation operators respectively.
The   indexes $\alpha$ and $\beta$ label each  of the three
levels, whereas the  index $\mu$ runs over all the atoms of type
$m$. We can introduce different type of atoms (let us say $L$
different types), each type of atom  associated with a copy of the
SU(3) algebra ($m=1,...,L$).

The operators $\mathbf{K}_{\alpha\beta m}$ have a simple physical
interpretation. For $\alpha$=$\beta$ the operators are  the number
operators: $\mathbf{K}_{\alpha\alpha m}=\mathbf{n}_{\alpha m}$.
The raising operators ($\alpha<\beta$) take a particle
(excitation) from a  level to a lower one, whereas the lowering
operators ($\alpha>\beta$) take a particle from a  level to a
higher one. In this context, the condition (\ref{conserva}) is simply the
conservation of the number of particles in each three level atom.

The linear terms appearing in the integrals of motion
(\ref{integsu3}) and (\ref{integsu3tr}) are related to the
energies of the atoms' levels, whereas the quadratic terms
represent two kinds of interactions: (a) the terms with
$\alpha\not= \beta$ include  three different dipole-dipole
interactions among the atoms, which are  associated, respectively,
with the transitions $1\leftrightarrow 2$, $1\leftrightarrow 3$ and
$ 2\leftrightarrow 3$ (see figure 1),  (b) the terms with
$\alpha$=$\beta$ are monopole interactions ($\mathbf{n}_{\alpha m
}\mathbf{n}_{\alpha n}$).
\begin{figure}
\includegraphics[width=6.3cm,height=5.2cm]{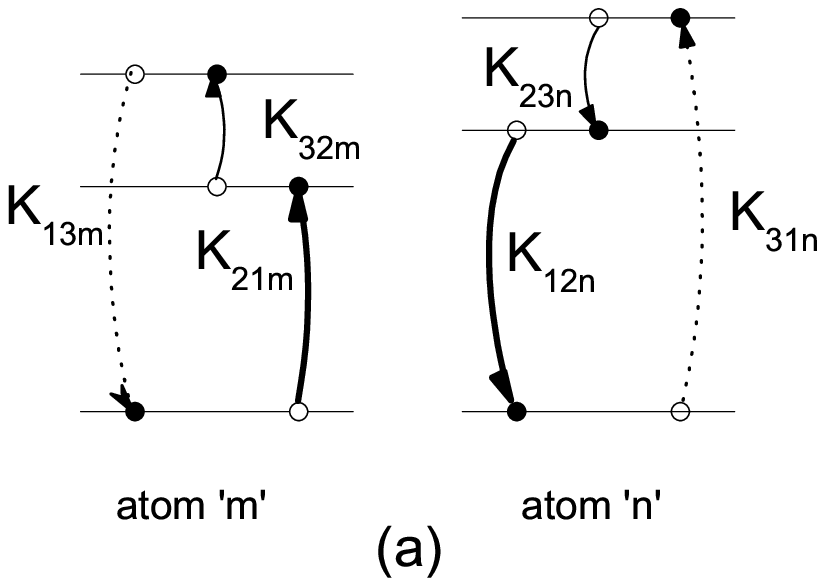}
\includegraphics[width=6.3cm,height=5cm]{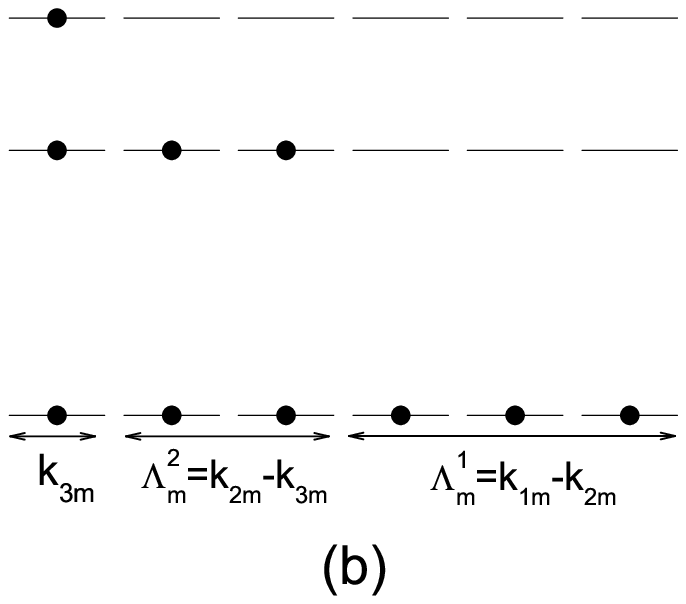}
\caption{(a) Dipole-dipole transitions between atoms of type $m$
and $n$. (b) Pictorial representation of a highest weight state
for  atoms of type $m$. }
\end{figure}

We discuss now the meaning of the labels that characterize the
SU(3) irreps (the set $(\Lambda_m^1,\Lambda_m^2)$ and $n_m$, or
$[k_{1m},k_{2m},k_{3m}]$). Without loss of generality, we can
consider $k_{1 m}=\Omega_m$ ($\Omega_m-k_{1 m}$ is  the number of
atoms without any particle in their  levels, which  are completely
decoupled of the rest and do not contribute to the $r_m$
eigenvalues).  The atoms of  each type  can be classified
according to the number of particle in their levels: (a) {\it
unblocked atoms} are those with only a particle in their levels.
These atoms have all the dipole transition allowed. (b) {\it
semi-blocked atoms} are those with two particles in their levels,
the dipole transitions between the occupied levels are forbidden
by Pauli blocking. (c) {\it blocked atoms } are those with all
their levels occupied, these atoms have all the dipole transitions
forbidden, and interact with the others only by the monopole
interaction. The Dynkin labels of the irreps determine the number
of atoms of each type  in the previous situations:
$\Lambda_m^1=k_{1m}-k_{2m}\leq \Omega_m$ indicates the number of
type-$m$ {\it unblocked  atoms}, $\Lambda_m^2=k_{2m}- k_{3m}$ is the
number of  {\it semi-blocked atoms} of type $m$, and  the value
$k_{3m}$ indicates the number of {\it blocked atoms} of type $m$.
A  representation of a highest weight state is depicted
in figure 1. The case where all the  atoms are  in a {\it
unblocked} situation corresponds to $\Lambda_m^1=\Omega_m$,
$\Lambda_m^2=0$ and $k_{3m}=0$.  Note that integrals of motion
(\ref{noverall}) guarantee that the  total populations
(irrespective on the type of atom) in each level ($N_\alpha,\
\alpha=1,2,3$) are conserved quantities.
 The number of parameters $E_k$ and $\omega_l$ in
the Richardson-Bethe equations (\ref{limitesu3}) has a simple
meaning: the number of $E_k$`s ($M_1$) is equal to the number of
atoms (of any type) with the first level unoccupied, whereas the
number of $\omega_l$`s ($M_2$) is the overall number of unblocked
and semi-blocked  atoms with the third level occupied.

\subsection{Particle-particle representation}

Another  possible realization of the SU(3) algebra 
is:
 \begin{eqnarray}
\mathbf{K}_{21 m}=\mathbf{c}_{\downarrow m}^{ \dagger}
\mathbf{c}_{0 m}^\dagger\qquad& \mathbf{K}_{12 m}=\mathbf{c}_{0 m}
\mathbf{c}_{\downarrow m}\qquad&\mathbf{K}_{11
m}=1-\mathbf{n}_{0 m}\nonumber\\
\mathbf{K}_{31 m}=\mathbf{c}_{\uparrow m}^{ \dagger} \mathbf{c}_{0
m}^\dagger\qquad&\mathbf{K}_{13 m}=\mathbf{c}_{0 m}
\mathbf{c}_{\uparrow m}\qquad&   \mathbf{K}_{22
m}=\mathbf{n}_{\downarrow
m}\nonumber\\
\mathbf{K}_{32 m}=\mathbf{c}_{\uparrow m}^{ \dagger}
\mathbf{c}_{\downarrow m}\qquad& \mathbf{K}_{23
m}=\mathbf{c}_{\downarrow m}^{ \dagger} \mathbf{c}_{\uparrow
m}\qquad& \mathbf{K}_{33 m}=\mathbf{n}_{\uparrow m},
\label{operators2}
 \end{eqnarray}
 where $\mathbf{c}_{\sigma m}^\dagger$ and
$\mathbf{c}_{\sigma m}$ are, respectively, fermionic creation and
annihilation operators, and $\mathbf{n}_{\sigma m}$ are the number
operators $\mathbf{c}_{\sigma m}^\dagger\mathbf{c}_{\sigma m}$.
The two lowering operators $\mathbf{K}_{21}$ and $\mathbf{K}_{31}$
create a pair of particles and thus form a {\it spinorial} pair:
 \begin{equation}
\mathbf{P}_m=\left(\begin{array}{c} \mathbf{P}_{\uparrow m}^\dagger\\
\\ \mathbf{P}_{\downarrow m}^\dagger\\
\end{array}\right)\equiv\left(\begin{array}{c}  \mathbf{K}_{31 m}\\
\\
\mathbf{K}_{21 m}\end{array}\right). \end{equation} The other
lowering operator ($\mathbf{K}_{32 m}\equiv\mathbf{T}_{+ m}$) is
the ladder operator of a $SU_T(2)$ subalgebra of SU(3):
\begin{equation}[\mathbf{T}_{+ m}, \mathbf{P}_{\downarrow
m}^\dagger]=\mathbf{P}_{\uparrow m}^\dagger . \end{equation}
 It is well known that the  particle-particle  representation of
the RG models associated with the SU(2) algebra  describes the
interaction between  scalar pairs \cite{Duke2},  whereas the SO(5)
version describes the vectorial pairing, i.e., the interaction
between pairs forming a triplet \cite{Duke6}. The SU(3) version
presented here is associated with an interaction between pairs
forming a doublet, we call it {\it spinorial} pairing or $T=1/2$
pairing. As it was done in \cite{Duke6} for the SO(5) algebra,
from a linear combination of the rational integrals of motion
(\ref{integsu3}), the following pairing Hamiltonian for spinorial
pairs  results:
 \begin{equation}
\mathbf{H}_{T=1/2}=\sum_m  z_m \left(\mathbf{n}_{\uparrow m}+
\mathbf{n}_{\downarrow m}\right)+ g\sum_{m n}
\mathbf{P}_m^\dagger\cdot\mathbf{P}_{n}, \end{equation} where the
scalar product  is
$\sum_{\sigma=\uparrow\downarrow}\mathbf{P}_{\sigma
m}^\dagger\mathbf{P}_{\sigma n}$,  and we have fixed  the
parameters $\xi^1=1/g$ and $\xi^2=0$. In this particle-particle
representation of the SU(3) algebra,  the Dynkin labels of the
highest weight states   are associated with the number of unpaired
particles in each single particle level $m$ and their
transformation properties under the subgroup $SU_T(2)$. The number
of variables $E_k$ and $\omega_l$ is, respectively, the number of
pairs and the number of $\uparrow$-particles  in those pairs.

\subsection{Tavis-Cummings models for
three level systems and two bosons}

In this section we  derive another integrable model, which
consists of $L$  copies of the SU(3) algebra interacting with two
different bosons. This model is derived from the trigonometric
version of the previous section in two steps: first we  take the
bosonic mapping  of one copy of the SU(3) algebra (let us say the
copy $0$), and then we let the degeneracy $\Omega_0$ of this
bosonized copy  go to infinite.

We  assume the SU(3) irrep
$(\Lambda_0^1,\Lambda_0^2)=(\Omega_0,0)$ for the copy to be
bosonized. In this case the bosonic mapping for the SU(3) algebra
is \cite{klein}:

\begin{eqnarray}
\fl
 \mathbf{K}_{11}\rightarrow \Omega_0-\mathbf{a}^\dagger
\mathbf{a}-\mathbf{b}^\dagger \mathbf{b}\qquad &
\mathbf{K}_{22}\rightarrow \mathbf{a}^\dagger \mathbf{a}\qquad &
\mathbf{K}_{33}\rightarrow \mathbf{b}^\dagger \mathbf{b}  \\
\fl
\mathbf{K}_{31}\rightarrow\mathbf{b}^\dagger\sqrt{\Omega_0-\mathbf{a}^\dagger
\mathbf{a}-\mathbf{b}^\dagger\mathbf{b}} \qquad&\mathbf{K}_{32}\rightarrow \mathbf{ b}^\dagger \mathbf{a}\qquad& 
\mathbf{K}_{21}\rightarrow
\mathbf{a}^\dagger\sqrt{\Omega_0-\mathbf{a}^\dagger
\mathbf{a}-\mathbf{b}^\dagger\mathbf{b}}, \nonumber
 \label{mapklein}
\end{eqnarray}
 where   $\mathbf{a}^\dagger$ and $\mathbf{b}^\dagger$ are boson
operators. This mapping is hermitic, therefore the mapping
of the operators $\mathbf{K}_{\alpha\beta}$ with $\beta>\alpha$
can be obtained by conjugating the previous ones.
In  the limit $\Omega_0\rightarrow\infty$ the previous mapping is reduced to:
\begin{eqnarray}
\fl
 \mathbf{K}_{11}\rightarrow \Omega_0-\mathbf{a}^\dagger
\mathbf{a}-\mathbf{b}^\dagger \mathbf{b}\qquad &
\mathbf{K}_{22}\rightarrow \mathbf{a}^\dagger \mathbf{a}\qquad &
\mathbf{K}_{33}\rightarrow \mathbf{b}^\dagger \mathbf{b} \nonumber \\
\fl
\mathbf{K}_{31}\rightarrow \mathbf{i}\sqrt{\Omega_0}
\mathbf{ b}^\dagger\qquad& \mathbf{K}_{32}\rightarrow \mathbf{ b}^\dagger
\mathbf{a}\qquad&  \mathbf{K}_{21}\rightarrow
\mathbf{i}\sqrt{\Omega_0} \mathbf{ a}^\dagger ,
 \label{mapbos}\end{eqnarray}
where  the imaginary factor $\mathbf{i}$ is introduced
for future convenience.
We consider now the trigonometric integrals of the RG model
(\ref{integsu3tr}) for a system of  $L+1$ copies of the Lie
Algebra: $m=0,1,...,L$. Then we bosonize the copy with the "0"
label using (\ref{mapbos}). In order to avoid divergences in taking  the limit
$\Omega_0\rightarrow\infty$,
 we  define a new set of variables $\epsilon_m$  by using the freedom we have to choose the parameters $z_m$. These new variables  are defined
 through:
 \begin{equation}\cot(z_m-z_0)=\frac{\epsilon_m}{\sqrt{\Omega_0}}\label{defa}\qquad m=1,...,L\end{equation}
  From this definition and  the trigonometric identity
  ($\cot(A-B)=(\cot(A)\cot(B)+1)/(\cot(B)-\cot(A))$,
   it is easy to show that in the limit
 $\Omega_0\rightarrow \infty$:
\begin{equation}
\cot(z_m-z_n)\rightarrow\frac{\sqrt{\Omega_0}}{\epsilon_n-\epsilon_m}.
\label{defa2} \end{equation} By substituting these results in the
trigonometric integrals  $\mathbf{R}_0$ and
$\mathbf{R}_m$(\ref{integsu3tr}), we obtain:
 \begin{eqnarray}
 \fl
\hat\mathcal{R}_0=&-\zeta^1\mathbf{a}^\dagger
\mathbf{a}-(\zeta^1+\zeta^2)
 \mathbf{b}^\dagger \mathbf{b}+\mathcal{C}_0\mathbf{1}
\nonumber\\
\fl
 &+\sum_{n=1}^L
\left(\mathbf{a}^\dagger \mathbf{K}_{12 n}+\mathbf{a}
\mathbf{K}_{21 n}+ \mathbf{b} \mathbf{K}_{31 n}+\mathbf{b}^\dagger
\mathbf{K}_{13 n} +\epsilon_n \mathbf{K}_{11 n} \right)
 \label{bosinteg}\\
\fl
 \hat{\mathcal{R}}_m=& -\zeta^1\mathbf{K}_{22 m}
-(\zeta^1+\zeta^2)\mathbf{K}_{33
m}+\mathcal{C}_m\mathbf{1}+ \sum_{n(n\neq
m)}^L\frac{\sum_{\alpha\beta} \mathbf{K}_{\alpha\beta
m}\mathbf{K}_{\beta\alpha n} }{\epsilon_m-\epsilon_n}
 \nonumber\\
 \fl &-\left(\mathbf{a}^\dagger \mathbf{K}_{12 m}+\mathbf{a}
\mathbf{K}_{21 m}+ \mathbf{b} \mathbf{K}_{31 m}+\mathbf{b}^\dagger
\mathbf{K}_{13 m}
 +\epsilon_m \mathbf{K}_{11 m} \right) \nonumber,
\end{eqnarray} where we have rescaled the
operators $\mathbf{R}_0$ and $\mathbf{R}_m$
($\hat\mathcal{R}_0$ $\equiv$ $\mathbf{R}_0/\sqrt{\Omega_0}$ and
$\hat\mathcal{R}_m$ $\equiv$ $\mathbf{R}_m/\sqrt{\Omega_0}$), dropped the
terms with inverse powers of $\sqrt{\Omega_0}$, and defined new
variables:
\begin{equation}
\zeta^a=\frac{\xi^a}{\sqrt{\Omega_0}}.\end{equation}
The constants in the integrals of motion (\ref{bosinteg}) are
$\mathcal{C}_0=-(1/3)\sum_l \epsilon_l n_l+
(2\zeta^1+\zeta^2)(\Omega_0/3)$ and
$\mathcal{C}_m=(\epsilon_m n_m)/3+\sum_l'
n_l/(\epsilon_l-\epsilon_m) +(2\zeta^1+\zeta^2) n_m/3$.

 To  obtain the eigenvalues of the previous operators, we  take the eigenvalues
(\ref{evaluesu3}), with $Z(z_j-z_i)$ and $Z(E_k-z_i)$ substituted
by cotangent functions, and then  the limit
$\Omega_0\rightarrow\infty$. Before performing this limit we
introduce,  as we did in
(\ref{defa}),  two new sets of variables associated with the parameters $E_k$ and $\omega_k$: \begin{equation}
\cot(z_0-E_k)=-\frac{x_k}{\sqrt{\Omega_0}}\qquad
\cot(z_0-\omega_l)=-\frac{y_l}{\sqrt{\Omega_0}}.
 \end{equation}
 The same trigonometric identity that yields (\ref{defa2})  allows us to
 write the limit $\Omega_0\rightarrow\infty$ of the following
 functions:
 \begin{eqnarray}\fl
 \cot(z_m-E_k) \rightarrow \frac{\sqrt{\Omega_0}}{x_k-\epsilon_m}
\qquad \cot(z_m-\omega_l)\rightarrow \frac{\sqrt{\Omega_0}}{y_l-\epsilon_m}\qquad
 \cot(\omega_{l'}-\omega_l)\rightarrow\frac{\sqrt{\Omega_0}}{y_l-y_{l'}}\nonumber\\
\cot(E_k-E_l)\rightarrow \frac{\sqrt{\Omega_0}}{x_l-x_k}\qquad
\cot(E_k-\omega_l)\rightarrow \frac{\sqrt{\Omega_0}}{y_l-x_k},
 \label{newset}
\end{eqnarray} where we have used the definition (\ref{defa}).
 With  the limit values of the  previous cotangent functions, the eigenvalues (\ref{evaluesu3}) become:
 \begin{eqnarray}
\fl
\mathrm{r}_0
=&\sum_{n=1}^L
 k_{1 n }\epsilon_n+\mathcal{C}_0-\sum_{k=1}^{M_1}x_k\\
\fl
\mathrm{r}_m
 = &- \zeta^1 k_{2 m}
 -(\zeta^1+\zeta^2)k_{3 m}+
 \mathcal{C}_m+\sum_{n=1(n\neq m)}^L\frac{\sum_{\alpha=1}^3 k_{\alpha m}k_{\alpha n}}{\epsilon_m-\epsilon_n}
-k_{1 m}\epsilon_m \nonumber\\
\fl
 &+ (k_{1 m}-k_{2 m})\sum_{k=1}^{M_1}\frac{1}{x_k-\epsilon_m}
 +(k_{2 m}-k_{3 m})\sum_{l=1}^{M_2}\frac{1}{y_l-\epsilon_m}
 ,
\end{eqnarray}
which are the eigenvalues of the operators (\ref{bosinteg}). Note
that the term $\mathcal{C}_0$, that diverges in the limit
$\Omega_0\rightarrow\infty$,  appears explicitly both in the
operator and its eigenvalue, then we can easily get rid of it.
 $\epsilon_m$ are free parameters and the
variables $x_k$ and $y_l$ are solutions of the Richardson-Bethe
equations in the limit $\Omega_0\rightarrow\infty$. To derive
them, we consider the Richardson-Bethe equations in the
trigonometric case, which can be read from  equations
(\ref{richeqsu3-a}) and (\ref{richeqsu3-b}) with   $Z$ substituted
by cotangent functions. Then, using (\ref{newset}) we obtain:
\begin{eqnarray} \fl
  \sum_{k' (k'\neq
k)}^{M_1}\frac{2}{x_{k'}-x_{k}}-\sum_{l=1}^{M_2}\frac{1}{y_l-x_k}
-\sum_{m=1}^L\frac{k_{1 m}-k_{2
m}}{\epsilon_m-x_k}-x_k=-\zeta^1\quad &
(k=1,.., M_1)\nonumber \\
\fl
 -\sum_{k=1}^{M_1}\frac{1}{x_k-y_l}+\sum_{l' (l' \neq
l)}^{M_2}\frac{2}{y_{l'}-y_l} -\sum_{m=1}^L\frac{k_{2 m }-k_{3 m
}}{\epsilon_m-\omega_l}=-\zeta^2\quad & (l=1,.., M_2).
\label{richeqsbo}
\end{eqnarray}
  Note that the resulting Richardson-Bethe equations are identical
to those in the rational case except by the linear term in the
first line. The integrals of motion (\ref{noverall}) are now:
\begin{eqnarray} \mathbf{N}_{3}=\sum_{m=1}^L
\mathbf{K}_{33 m}+ \mathbf{n}_{b} \qquad  \mathbf{N}_{2}=
\sum_{m=1}^L\mathbf{K}_{22 m}+ \mathbf{n}_{a}\nonumber\\
\mathbf{N}_1= \sum_{m=1}^L \mathbf{K}_{11 m }+
\Omega_0-\mathbf{n}_a-\mathbf{n}_b.\end{eqnarray}
 The physical meaning of the
Dynking labels $(\Lambda_m^1,\Lambda_m^2)$ and $n_{m}$ is not
modified by the introduction of the bosons, and it is the same
already discussed above. The number of parameters $x_k$ and $y_l$
($M_1$ and $M_2$) in the Richardson-Bethe equations
(\ref{richeqsbo}) can be obtained from the general formula
(\ref{limitesu3}), by extending the sum to the bosonized copy, and
noting that the Young labels associated with it are
$[k_{10},k_{20},k_{30}]=[\Omega_0,0,0]$.

One of the simplest  Hamiltonian that can be derived from the
previous integrals of motion, is obtained by  considering just one
non-bosonized copy of the SU(3) algebra. Therefore $L=1$ and we
have 2 integrals of motion: $\mathbf{R}_0$ and $\mathbf{R}_1$. By
taking a linear combination of these integrals  we arrive to:
\begin{eqnarray} \fl \mathbf{H}&=&z_0 \left(\hat{\mathcal{R}}_0-\mathcal{C}_0\mathbf{1}\right) + z_1\left(
\hat{\mathcal{R}}_1-\mathcal{C}_1 \mathbf{1}\right)\nonumber\\
 \fl &= &\omega_a \mathbf{n}_a +
\omega_b \mathbf{n}_b + \eta_2 \mathbf{K}_{22}+
\eta_3\mathbf{K}_{33}+ g\left(\mathbf{a}^\dagger \mathbf{K}_{12
}+\mathbf{a} \mathbf{K}_{21 }+ \mathbf{b} \mathbf{K}_{31
}+\mathbf{b}^\dagger \mathbf{K}_{13 }
\right)\label{hamiltc},\end{eqnarray}
 where $\mathbf{K}_{\alpha\beta}$ are the generators of the non-bosonized copy.   The parameters of the Hamiltonian
are related to the RG parameters by: \begin{equation}\fl
\zeta^1=\frac{\omega_a(\Delta_{\eta}-\Delta_{\omega})}{g
\Delta_{\omega}}\qquad
\zeta^2=\frac{\Delta_{\eta}-\Delta_{\omega}}{g} \qquad
 \epsilon_1=\frac {\omega_a\Delta_{\eta} -\Delta_\omega\eta_2 }{g\Delta_{\omega}} , \end{equation} and the
constants $z_{0}$ and $z_1$ are: \begin{equation} z_0=-g
\frac{\Delta_{\omega}}{\Delta_{\eta}-\Delta_{\omega}}\qquad
z_1=-g\frac{\Delta_{\eta}}{\Delta_{\eta}-\Delta_{\omega}},
\end{equation} with $  \Delta_{\omega}\equiv \omega_b - \omega_a$,
and $ \Delta_{\eta}\equiv \eta_3 - \eta_2$. The previous
Hamiltonian describes the interaction between a set of $\Omega_1$  identical
three level atoms and two modes of the electric field in the so
called Rotating-Wave-Approximation \cite{TC3}, and is a
generalized version of the Tavis-Cummings model for three level
atoms.

 If we write the Hamiltonian (\ref{hamiltc}) in the
particle-particle representation, we obtain the interaction
between a doublet of pairs and a doublet of bosons:
\begin{equation} \mathbf{H}=\omega_+ \mathbf{n}_{+} + \omega_-
\mathbf{n}_{-} + \epsilon_\uparrow \mathbf{n}_{\uparrow}+
\epsilon_\downarrow\mathbf{n}_\downarrow+ g\left(
\mathbf{P}^\dagger\cdot\mathbf{b} +
\mathbf{P}\cdot\mathbf{b}^\dagger \
\right)\label{hamilpe},\end{equation} where we have redefined the
bosons $\mathbf{a}^\dagger\rightarrow \mathbf{b}_{-}^\dagger$,
$\mathbf{b}^\dagger\rightarrow \mathbf{b}_{+}^\dagger$, and the
scalar product,  $\mathbf{P}^ \dagger\cdot\mathbf{b}$, is $
\mathbf{P}^\dagger_{\uparrow}\mathbf{b}_{+}+\mathbf{P}^\dagger_{\downarrow}\mathbf{b}_{-}$.
More complex Hamiltonians can be derived from the integrals
(\ref{bosinteg}), but we will not discuss them here.

\section{Conclusions}
In this contribution we introduced a Gaudin algebra valid for any
simple Lie algebra in the rational,  trigonometric, and hyperbolic
cases. With  this algebra,  we derived the  complete set of
quadratic RG integrals of motion, and found their respective
eigenvalues. Focusing this  formalism  on the rank-two SU(3)
algebra, we worked out in detail the RG models. For the rational
case, expressions for the integrals of motion of degree three were
obtained, and it was verified that their number is the necessary
to match the number of integrals of motion with the number of
degrees of freedom of the quantum model. Some physical
applications of the SU(3) models were discussed. The physical
meaning of the variables in the RG formalism were discussed for
two different representations of the SU(3) algebra, namely, the
particle-hole representation and the particle-particle one. By
taking the trigonometric version of the models, we derived a new
family of integrable models which are related to a generalization
of the
Tavis-Cummings model to three level atoms and two bosons. 
It was out of the scope of this contribution to explore all the
possible physical applications of the SU(3) RG models, some
examples were given and others are expected to come. We think that
we have put the ground for more detailed studies, such as
numerical studies of the solutions and comparison with
approximative techniques in physical scenarios beyond the limits
of traditional diagonalization methods. The formalism presented
here can be easily applied to other Lie algebras. Some physically
relevant examples are the SU(4) and SO(5) versions in the
description of High temperature superconductors \cite{htsc}, the
SO(8) version for isovector-isoscalar  pairing \cite{so8pairing},
and the SU(n) versions which include generalized  Tavis-Cummings
models  for $n$-levels atoms interacting with $n-1$ different
bosonic modes. The formalism can likewise be extended to the
elliptic RG models, where no linear term in  the quadratic
integrals of motion  is allowed.  Another interesting extension of
the present contribution is to derive more general solutions to
the Gaudin conditions (\ref{gaud1}), using the very well studied
solutions to the Classical Yang-Baxter Equations \cite{cybesol}.
Finally, the higher rank RG models can be useful to shed some
light in the non completely well established definition of number
of degrees of freedom in finite Quantum systems. In this
contribution, the definition of integrability coming from the
Yang-Baxter Equation was linked to the one  coming from the
equality between the number of quantum degrees of freedom and the
number of independent integrals of motion. This latter definition,
initially supposed to be exclusively related to dynamical
symmetric models,  can be  extended, as shown in this
contribution,  to the RG models, which are not necessarily
dynamical symmetric.

\ack This work was supported in part by the Spanish DGI under
grant No. BFM2003-05316-C02-02. B. Errea has  a pre-doctoral grant from CE-Comunidad Aut\'onoma de
Madrid and  S. Lerma H.  a  post-doctoral
grant from Spanish SEUI-Ministerio de Educaci\'on y Ciencia. The authors thank J. Dukelsky, S. Pittel and P. Van
Isacker for useful discussions and suggestions to the manuscript.
S L H wishes to express his sincere thanks to J. Dukelsky for suggesting and introducing him in  this subject.


\section*{References}

\begin{thebibliography}{99}

\bibitem{Duke1} Dukelsky J, Pittel S and Sierra G 2004 {\it Rev.\ Mod.\
Phys.}
\textbf{76} 643

\bibitem{Links3} Links J, Zhou H-Q, McKenzie R H and Gould M D 2003 {\it J.\
Phys.}\ A \textbf{36} R63-R104
\bibitem{Rich1}
Richardson R W 1963 {\it Phys.\  Lett.} \textbf{3}  277; Richardson
R W 1966 {\it Phys. Rev.} {\bf 141} 949


\bibitem{Gaudin1}  Gaudin M 1976 {\it J.\  Phys. (Paris)} {\bf 37} 1087

\bibitem{poghoss} von Delft J and  Poghossian R 2002 {\it Phys.\  Rev.}
B \textbf{66} 134502

\bibitem{Jurco} Jurco B 1989  {\it J. Math. Phys.} {\bf 30} 1289

\bibitem{Links2} Guan X-W,  Foerster A,  Links J and Zhou H-Q 2002 {\it Nucl.\
Phys.} B {\bf 642} 501

\bibitem{Sierra} Asorey M, Falceto F and Sierra G 2002 {\it Nucl.\  Phys.} B \textbf{622}
593


\bibitem{Ush} Ushveridze A G 1994 {\it Quasi-exactly solvable models in quantum
mechanics} (Bristol: Institute of Physics) p~357

\bibitem{Ortiz}  Ortiz G,  Somma R,  Dukelsky J and  Rombouts S 2005  {\it Nucl.\  Phys. } B {\bf 707} 421

\bibitem{Ami} Amico L,  Di Lorenzo A and  Osterloch A 2001 {\it Phys.\  Rev.\
Lett.} {\bf 86} 5759

\bibitem{Duke2} Dukelsky J, Esebbag C and Schuck P 2001 {\it Phys.\  Rev.\
Lett.} \textbf{87} 066403

\bibitem{Duke3} Dukelsky J, Dussel G G, Esebbag C and Pittel S 2004 {\it Phys.\  Rev.\  Lett.} {\bf 93} 050403


\bibitem{Links1}
Links J, Zhou H-Q, Gould M D and McKenzie R H 2002 {\it J.\
Phys.}\ A \textbf{35} 6459

\bibitem{Duke6}  Dukelsky J,  Gueorguiev V,  Van Isacker P,
Dimitrova S,  Errea B and  Lerma H S 2006 {\it Phys.\  Rev.\  Lett.}
 {\bf 96} 072503

\bibitem{Lerma}  Lerma H S,  Errea B,  Dukelsky J,  Pittel S and
 Van Isacker P 2006 {\it Phys.\  Rev} C {\bf 74} 024314

\bibitem{TC3}  Yoo H I and Eberly J H 1985 {\it Phys.\  Rep.} {\bf
118} 239

\bibitem{Talal1} Chervov A, Rybnikov L and Talalaev D 2004 Rational Lax operators and their quantization {\it Preprint} hep-th/0404106

\bibitem{Talal2}  Talalaev D 2006 {\it Funct.\  Anal.\  Appl.} {\bf 40}  73

\bibitem{Zhang} Zhang W M and  Feng D H 1995 {\it Phys.\ Rep.}
\textbf{252} 1



\bibitem{CFThe} Di Francesco P,  Mathieu P and  S\'en\'echal D 1997 {\it Conformal Field
Theory} (New York: Springer) p~489

\bibitem{Wyb74} Wybourne B G 1974 {\it Classical Groups for Physicists}
(New York: Wiley-Interscience)

\bibitem{klein}
 Klein A and  Marshalek E R 1991 {\it Rev.\ Mod.\ Phys.} {\bf 63} 2

\bibitem{htsc}  Guidry M,  Wu L-A,  Sun Y and  Wu C-L  2001 {\it Phys.\  Rev.} B {\bf 63} 134516

\bibitem{so8pairing}  Evans J A,  Dussel G G,  Maqueda E E and
 Perazzo P J 1981 {\it Nucl.\  Phys.} A {\bf 367} 77

\bibitem{cybesol}  Belavin A A and  Drinfel'd V G   1982 {\it Funct.\  Anal.\
Appl.} {\bf 16} 159

\end{thebibliography}
\end{document}